\documentclass[12pt]{iopart}

\usepackage{graphicx}
\usepackage{bm}

\begin{document}

\title[Cubic oscillator]{Global solution of the cubic oscillator}

\author{E M Ferreira$^1$ and J Sesma$^2$}
\address{$^1$ Instituto de F\'{\i}sica, Universidade Federal do Rio
de Janeiro, 21941-972, Rio de Janeiro, Brasil}
\address{$^2$ Departamento de F\'{\i}sica Te\'orica, Facultad de
Ciencias, 50009, Zaragoza, Spain}

\eads{\mailto{erasmo@if.ufrj.br}, \mailto{javier@unizar.es}}

\begin{abstract}
The eigenstates of a real or complex cubic anharmonic oscillator are investigated using an original and alternative method.
The procedure consists of determining global solutions of the Schr\"odinger equation that comply with the pertinent boundary
conditions   and allows us to obtain, in a very simple way, the eigenenergies and eigenfunctions of the Hamiltonian.
Scattering by a real cubic potential is investigated as a particular case.
\end{abstract}



\section{Introduction}

The cubic anharmonic oscillator is an interesting example of quantum mechanical system that cannot be solved by a conventional use of perturbation theory \cite[chapter VII, section 48]{davi}. A heuristic application of the dilatation-transformation method allowed Yaris {\it et al} \cite{yari} to identify localized solutions of the cubic anharmonic oscillator that could be interpreted as resonances. Caliceti, Graffi, and Maioli's rigorous study \cite{cali} of this problem with a complex coupling constant revealed the existence of a discrete spectrum, which admits analytic continuation as the imaginary part of the coupling constant goes to zero. They concluded that one can speak of resonances in a real cubic anharmonic oscillator. The behavior of these resonances for small and large values of the coupling constant was discussed by Alvarez \cite{alva1}, who also showed, in a different paper \cite{alva2}, that the quantum Rayleigh-Schr\"odinger perturbation series can be rearranged as the classical Birkhoff series plus quantum corrections in correspondence with the Jeffreys-Wentzel-Kramers-Brillouin (JWKB) series. Investigations on different aspects of the cubic anharmonic oscillator, like Bender-Wu branch points \cite{alva3}, JWKB expansions \cite{alva4}, and  precise determination of eigenvalues \cite{alva5,jent}, have continued up to the present.

Additional interest in cubic potential arose from the conjecture of Bessis and Zinn-Justin about the positivity and reality of its spectrum in the case of pure imaginary coupling constant. This observation motivated the birth of the $\mathcal{PT}$-symmetric quantum theory \cite{bend1}, a very fertile field of research  where the cubic anharmonic oscillator has been
subject of several studies   \cite{fern}. (For a recent explanation of the foundations of this theory see \cite{bend2}.)

The interest stirred by the cubic oscillator has reached potentials of the form $x^N$, with integer $N$. Voros's paper \cite{voro} deserves a special mention, as it uses very elegant methods of functional analysis to establish a relation between the cases of even and odd $N$. In particular, the author proves a duality between the linear ($N=1$) an the quartic ($N=4$) potentials which allow him to derive, as a byproduct, new properties of the well known Airy function and its zeros. Zinn-Justin and collaborators, in a series of papers \cite{zin1}, developed a procedure, based on instanton configurations, to be rid of the divergence of the perturbation series, a disease of anharmonic oscillators, and to obtain, for the eigenvalues, a resurgent expansion in terms of the coupling constant. The procedure has been applied in the computation of the eigenvalues of the cubic oscillator \cite{jent}. Besides, it allows one to establish a bridge between even anharmonic oscillators with negative coupling and odd anharmonic oscillators with imaginary coupling (i. e., $\mathcal{PT}$-symmetric ones) leading to a unified treatment of anharmonic oscillators of both parities \cite{prl}.

The analytic properties of the eigenvalues of the cubic oscillator, considered as a function of the intensity (assumed to be complex) of the cubic term, have been discussed recently. Two papers deal with the (divergent) perturbation series for the eigenvalues summed by means of order-dependent mappings \cite{zinn} or Pad\'e approximants \cite{grec}. A different strategy has been adopted in a paper by Fern\'andez and Garc\'{\i}a \cite{fega}, who use both complex rotation and Riccati-Pad\'e methods to determine the eigenvalues of the Hamiltonian with  extraordinary accuracy.

In this paper, we are concerned  with the study of the eigenstates of a cubic anharmonic oscillator with a complex coupling parameter.
Our approach to the problem, however, is a rigorous one: we obtain a global solution of the Schr\"odinger equation that satisfies the appropriate boundary conditions. The general problem of finding global solutions of a second-order differential equation with one or two singular points was solved, forty years ago, by Naundorf \cite{naun}. An improvement of his method has been proposed \cite{gom1} and applied in the study of several quantum problems \cite{gom2}. The procedure consists of writing exact solutions of the Schr\"odinger equation, in the form of convergent series expansions, and analyzing their behavior far from the origin. The requirement that this  behavior be regular determines the eigenvalues. In addition, the procedure provides algebraic expressions of the eigenfunctions, which facilitate easy computation of the expected value of any quantity.

Different notations have been used to represent the Hamiltonian of the cubic anharmonic oscillator. We adopt here the notation of \cite{yari}, namely
\begin{equation}
H(\kappa, \lambda) = -\,\frac{{\rm d}^2}{{\rm d}x^2}+\kappa\,\frac{x^2}{4}-\lambda\,x^3,  \label{uno}
\end{equation}
where $\kappa$ is assumed to be real, while $\lambda$ may be complex. Notice that we have introduced an additional parameter, $\kappa$, which is redundant: An adequate scaling of the variable $x$ would transform the Hamiltonian (\ref{uno}) into another one with a fixed value $\kappa=1$, for instance, and a scaled value of $\lambda$. Nevertheless, we prefer to maintain the Hamiltonian in the form (\ref{uno}) to facilitate the discussion of the interesting case of $\kappa=0$. Concerning the parameter $\lambda$, it is sufficient to consider its values in the sector $0\leq \arg \lambda\leq \pi/2$. Eigenvalues and eigenfunctions of $H$ with $\lambda$ in any other sector could be trivially obtained from those for $\lambda$ in the first quadrant by complex conjugation and/or reflection $x\longrightarrow -x$.

We deal with the solutions of the Schr\"odinger equation corresponding to the Hamiltonian (\ref{uno}) in section 2.  Two kinds of solutions are considered: series expansions, convergent in the finite complex $x$-plane, and formal asymptotic solutions, with a well defined behavior for $x\to\infty$. The connection between the two kinds of solutions, which is crucial point of our method for
the determination of eigenvalues, is discussed in section 3. The particular case of real $\lambda$ is treated in section 4, where the scattering problem is formulated. Sections 5 and 6 deal with the eigenvalues and eigenfunctions  of the Hamiltonian
and show numerical results for certain values of the parameters. The phase shift and the time delay in the scattering by a cubic real potential are illustrated in section 7. Some final comments are added in section 8.

\section{Solutions of the Schr\"odinger equation}

Finding the eigenvalues and eigenfunctions of the Hamiltonian given in Eq. (\ref{uno}) is a quantum mechanics exercise similar to that of determining bound states in a harmonic oscillator or in a Coulomb potential. The solution of the Schr\"odinger equation, for undetermined energy, is written as a power series expansion around the origin. The coefficients of the expansion are obtained by following the well known method of solving a second-degree homogeneous differential equation. Since, in the mentioned problems, the nearest-to-the-origin singular point of the Schr\"odinger equation is at infinity, the convergence of the series is guaranteed for finite values of the variable. At infinity, however, the series becomes divergent unless the energy takes certain discrete values. There is, nevertheless, a difference between the aforementioned algebraically solvable problems and the problem at hand. In those cases, the recurrence relation obeyed by the coefficients of the power series solution is a difference equation (of the hypergeometric class) that can be solved algebraically, making possible to determine the behavior of the power series solution at infinity. In the cubic oscillator case, instead, the rank of the singularity at infinity is larger. The difference equation that gives the coefficients of the power series solution is of higher order (it does not belong to the hypergeometric class), so it cannot be solved algebraically and, as a result, the behavior at infinity of the power series solution cannot be obtained so easily. The procedure to be followed here consists of considering, aside from and independently of the conventional (Frobenius) power series around the origin, asymptotic (Thom\'{e}) solutions in the vicinity of $-\infty$ and $+\infty$, selecting those being physically acceptable (regular) at the singular point. The next step is to require the coincidence, up to a constant factor, of the power series (Frobenius) solution with the asymptotic (Thom\'{e}) solutions regular at $-\infty$ and $+\infty$. Both conditions can be satisfied only for certain values of the energy, which can be determined in this way.

Let us use $E$ to denote the eigenvalues of the Hamiltonian, and $\psi (x)$ to denote the eigenfunctions. These satisfy the Schr\"odinger equation
\begin{equation}
-\,\frac{{\rm d}^2\psi(x)}{{\rm d}x^2}+\left(\kappa\,\frac{x^2}{4}-\lambda\,x^3-E\right)\psi(x)=0\,,  \label{dos}
\end{equation}
with adequate boundary conditions to be given below. Instead of solving the Schr\"odinger equation on the whole real axis, we may solve it only on the positive semiaxis with a signal, $\sigma$, indicating whether we are considering the true eigenfunction or its mirror image with respect to the vertical axis, $x=0$. Then, we are led to the differential equation
\begin{equation}
\fl \frac{{\rm d}^2\phi^{(\sigma)}(x)}{{\rm d}x^2}+\left(\sigma\lambda\,x^3-\frac{\kappa}{4}\,x^2+E\right)\phi^{(\sigma)}(x)=0\,, \qquad 0\leq x <+\infty \,, \quad \sigma=\pm 1\,,    \label{dosbis}
\end{equation}
where
\begin{equation}
\phi^{(-1)}(x)=\psi(-x)\,, \qquad \phi^{(+1)}(x)=\psi(x)\,, \qquad 0\leq x <+\infty \,, \label{dostri}
\end{equation}

To avoid the presence  of series of non-integer powers, we introduce a new variable,
\begin{equation}
t=x^{1/2}.   \label{tres}
\end{equation}
The function
\begin{equation}
w^{(\sigma)}(t)= x^{-1/4}\,\phi^{(\sigma)}(x)  \label{cuatro}
\end{equation}
obeys the differential equation
\begin{equation}
\fl t^2\,\frac{{\rm d}^2w^{(\sigma)}(t)}{{\rm d}t^2}+\left(4\sigma\lambda\,t^{10}-\kappa\,t^8+4E\,t^4-\frac{3}{4}\right)w^{(\sigma)}(t)=0, \qquad 0\leq t<+\infty\,.  \label{cinco}
\end{equation}
Two independent solutions of this equation can be written as series expansions (Frobenius solutions),
\begin{equation}
w_i^{(\sigma)}(t) = t^{\nu_i}\,\sum_{n=0}^\infty c_{n,i}^{(\sigma)}\,t^n, \qquad c_{0,i}^{(\sigma)}\neq 0, \qquad i=1, 2,  \label{seis}
\end{equation}
with indices
\begin{equation}
\nu_1=-\,1/2, \qquad \nu_2=3/2,  \label{siete}
\end{equation}
and coefficients given by the recurrence relation
\begin{equation}
\fl c_{0,i}^{(\sigma)}=1, \qquad c_{2,1}^{(\sigma)}=0\,,   \qquad  c_{n,i}^{(\sigma)}=\frac{-4E\,c_{n-4,i}^{(\sigma)}+\kappa\,c_{n-8,i}^{(\sigma)}-4\sigma\lambda\, c_{n-10,i}^{(\sigma)}}{n(n-1+2\nu_i)}.  \label{ocho}
\end{equation}
Any solution of Eq. (\ref{cinco}) may be written as a linear combination, and in particular, if we denote by $w_{\rm phys}$
the solution satisfying the adequate boundary conditions, to be specified below, we can write
\begin{equation}
 w_{\rm phys}^{(\sigma)}(t) \propto A_1^{(\sigma)}\,w_1^{(\sigma)}(t)+A_2^{(\sigma)}\,w_2^{(\sigma)}(t),  \label{nueve}
\end{equation}
with constants $A_1^{(\sigma)}$ and $A_2^{(\sigma)}$ to be determined. The continuity of $\psi_{\rm phys}(x)$ and of its derivative at $x=0$ is guaranteed if we
take
\begin{equation}
A_1^{(-1)}=A_1^{(+1)}=A_1\,, \qquad  A_2^{(-1)}=-\,A_2^{(+1)}=-\,A_2\,.  \label{nuevebis}
\end{equation}

Formal solutions in terms of asymptotic expansions for $t\to\infty$ (Thom\'e solutions) are also easily obtained. They are
\begin{equation}
\fl w_j^{(\sigma)}(t) = \exp\left(\alpha_j^{(\sigma)}\,t^5/5+\beta_j^{(\sigma)}\,t^3/3+\gamma_j^{(\sigma)}\,t\right)\,t^{-2}\,\sum_{m=0}^\infty a_{m,j}^{(\sigma)}\,t^{-m}, \quad a_{0,j}^{(\sigma)}\neq 0, \quad j=3, 4,  \label{diez}
\end{equation}
with exponents
\begin{equation}
\fl \alpha_j^{(\sigma)}=2\,(-\sigma\lambda)^{1/2}, \qquad \beta_j^{(\sigma)}=\kappa/(2\alpha_j^{(\sigma)}), \qquad
\gamma_j^{(\sigma)}=-\left(\beta_j^{(\sigma)}\right)^2/(2\alpha_j^{(\sigma)})\,, \label{duno}
\end{equation}
where the values of the subindex, $j$, are associated to the two possible values of $(-\sigma\lambda)^{1/2}$. We assign the
subindices in such a way that, for $0\leq\arg \lambda\leq\pi/2$, we have
\begin{eqnarray}
-\,\pi&\leq \arg \alpha_3^{(-1)}\leq-\,\frac{3\pi}{4}\,, \qquad \, 0\leq\arg \alpha_4^{(-1)}\leq\frac{\pi}{4}\,,  \nonumber  \\
\hspace{12pt}\frac{\pi}{2}&\leq \arg \alpha_3^{(+1)}\leq\frac{3\pi}{4}\,, \qquad  -\,\frac{\pi}{2}\leq\arg \alpha_4^{(+1)}\leq-\,\frac{\pi}{4}\,.  \nonumber
\end{eqnarray}
Accordingly, $w_3^{(-1)}(t)$ vanishes as $t\to\infty$, whereas $w_4^{(-1)}(t)$ becomes divergent. In a similar way, $w_3^{(+1)}(t)$ vanishes and $w_4^{(+1)}(t)$ diverges, for $t\to\infty$, whenever $\arg \lambda >0$. For real $\lambda$, both $\alpha_3^{(+1)}$ and $\alpha_4^{(+1)}$ are purely imaginary and the functions $w_3^{(+1)}(t)$ and $w_4^{(+1)}(t)$ oscillate for $t\to\infty$. They correspond to outgoing (to the right) and incoming (from the right) waves, respectively.
The coefficients, $a_{m,j}^{(\sigma)}$, of the asymptotic expansions in Eq. (\ref{diez}) obey the recurrence relation (omitting the superindex $(\sigma)$ and the subindex $j$)
\begin{eqnarray}
\fl a_0=1,  \qquad  2\,m\,\alpha\,a_{m}&=&(4E+2\beta\gamma)\,a_{m-1}-2(m-1)\beta\,a_{m-2}+\gamma^2\,a_{m-3}  \nonumber \\
& &-\,2(m-2)\gamma\,a_{m-4}+((m-2)(m-3)-3/4)\,a_{m-5}.  \label{ddos}
\end{eqnarray}

The behavior of the Frobenius solutions at infinity can be written in terms of the Thom\'e solutions by means of the connection factors, $T$. These are independent of $t$, but depend on the values of the parameters, $\kappa$ and $\lambda$, and on the energy, $E$. We have,
for $t\to\infty$,
\begin{equation}
w_i^{(\sigma)}(t) \sim  T_{i,3}^{(\sigma)}\,w_3^{(\sigma)}(t)+T_{i,4}^{(\sigma)}\,w_4^{(\sigma)}(t),\qquad i=1, 2. \label{dtres} \\
\end{equation}
(Notice that the connection factors vary from one sector of the complex plane to others, a fact that is known as Stokes phenomenon.) From the above discussion about the behavior of the Thom\'e solutions for $t\to\infty$, it follows that the regularity of $w_{\rm phys}^{(-1)}$ requires the coefficients $A_1$ and $A_2$ to be chosen in such a way that
\begin{equation}
A_1\,T_{1,4}^{(-1)}-A_2\,T_{2,4}^{(-1)}=0\,.  \label{extra1}
\end{equation}
Analogously, for $\arg \lambda >0$, the regularity of $w_{\rm phys}^{(+1)}$ requires that $A_1$ and $A_2$ satisfy
\begin{equation}
A_1\,T_{1,4}^{(+1)}+A_2\,T_{2,4}^{(+1)}=0\,.  \label{extra2}
\end{equation}
(The special case of real $\lambda$, i. e. when $w_3^{(+1)}(t)$ and $w_4^{(+1)}(t)$ are oscillating waves for $t\to\infty$, will be discussed below.)
Both conditions (\ref{extra1}) and (\ref{extra2}) can be fulfilled if and only if
\begin{equation}
T_{1,4}^{(+1)}\,T_{2,4}^{(-1)}+ T_{2,4}^{(+1)}\,T_{1,4}^{(-1)}=0.  \label{dseis}
\end{equation}
For given $\kappa$ and $\lambda$, the last equation is a restriction on the values of $E$ and, therefore, provides a procedure for determining the eigenvalues of the Hamiltonian (\ref{uno}), whenever a way for obtaining the connection factors, $T_{j,k}$, is available. This is the crucial point in our method for solving the Schr\"odinger equation.

\section{Determination of the connection factors}

By taking the Wronskian of the two sides of Eq. (\ref{dtres}) with $w_4^{(\sigma)}$ and $w_3^{(\sigma)}$, we obtain
\begin{equation}
T_{i,3}^{(\sigma)}=\frac{\mathcal{W}[w_i^{(\sigma)},w_4^{(\sigma)}]}{\mathcal{W}[w_3^{(\sigma)},w_4^{(\sigma)}]}, \qquad T_{i,4}^{(\sigma)}=\frac{\mathcal{W}[w_i^{(\sigma)},w_3^{(\sigma)}]}{\mathcal{W}[w_4^{(\sigma)},w_3^{(\sigma)}]},
 \qquad i=1, 2, \label{dsiete}
\end{equation}
where we have used the symbol $\mathcal{W}[f,g]$ to represent the Wronskian of the two functions $f(t)$ and $g(t)$, namely
\[
\mathcal{W}[f,g]\equiv f(t)\,\frac{{\rm d}g(t)}{{\rm d}t}-\frac{{\rm d}f(t)}{{\rm d}t}\,g(t).
\]
Notice that all Wronskians in the right-hand  side of Eq. (\ref{dsiete}) are independent of $t$, since the involved functions are solutions of the same
second-order homogeneous differential equation in which the first derivative term is absent. The Wronskians in the denominators can be computed
directly by using the expansions Eq. (\ref{diez}) to obtain
\begin{equation}
\mathcal{W}[w_3^{(\sigma)},w_4^{(\sigma)}] = - \,\mathcal{W}[w_4^{(\sigma)},w_3^{(\sigma)}] =  2\,\alpha_4^{(\sigma)}\,.  \label{docho}
\end{equation}
The evaluation of the Wronskians in the numerators is not so easy. In a previous paper \cite{gom1}, we provided a general procedure to calculate Wronskians of Frobenius and Thom\'e solutions of a Schr\"odinger equation with a polynomial potential. For the convenience of the reader, we reproduce the method here, adapted to the present case.

Our goal is to obtain $\mathcal{W}[w_i^{(\sigma)},w_j^{(\sigma)}]$. As a first step, we introduce auxiliary functions
\begin{eqnarray}
 \fl u_{i,j}^{(\sigma)}(t)=\exp\left( -\alpha_j^{(\sigma)}\,t^{5}/10\right)\,w_{i}^{(\sigma)}(t),& \qquad u_j^{(\sigma)}(t)=\exp\left( -\alpha_j^{(\sigma)}\,t^{5}/10\right)\,w_j^{(\sigma)}(t), \label{dnueve} \\
  &  i=1, 2,  \qquad  j=3, 4 ,   \nonumber
\end{eqnarray}
whose Wronskian is closely related to the one that we want to determine. In fact,
\begin{equation}
\mathcal{W}[u_{i,j}^{(\sigma)},u_j^{(\sigma)}]=\exp\left(-\,\alpha_j^{(\sigma)}\,t^{5}/5\right) \,\mathcal{W}[w_{i}^{(\sigma)},w_j^{(\sigma)}]. \label{veinte}
\end{equation}
We can write a formal expansion of the left-hand  side of this equation by using the definitions (\ref{dnueve}) and the expansion (\ref{diez}), obtaining
\begin{equation}
\fl \mathcal{W}[u_{i,j}^{(\sigma)},u_j^{(\sigma)}] \sim \left(\left(\alpha_j^{(\sigma)}\,t^{4}+2\left(\beta_j^{(\sigma)}\,t^{2}+\gamma_j^{(\sigma)}\right)\right)v_{i,j}^{(\sigma)}
-\frac{{\rm d}v_{i,j}^{(\sigma)}}{{\rm d}t}\right)\mathcal{S}_j^{(\sigma)}
+ v_{i,j}^{(\sigma)}\,\frac{{\rm d}\mathcal{S}_j^{(\sigma)}}{{\rm d}t},   \label{vuno}
\end{equation}
where we have denoted
\begin{equation}
\fl v_{i,j}^{(\sigma)}=\exp\left(\beta_j^{(\sigma)}\,t^3/3+\gamma_j^{(\sigma)}\,t\right)\,w_i^{(\sigma)}, \quad
\mathcal{S}_j^{(\sigma)} = \sum_{m=0}^\infty a_{m,j}^{(\sigma)}\,t^{-m-2}, \quad i=1, 2\,, \quad j=3, 4\,.  \label{vdos}
\end{equation}
A series expansion for $v_{i,j}^{(\sigma)}$,
\begin{equation}
v_{i,j}^{(\sigma)} = \sum_{n=0}^\infty \hat{c}_{n,i,j}^{(\sigma)}\,t^{n+\nu_i},     \label{vcuatro}
\end{equation}
can be obtained from its definition, Eq. (\ref{vdos}), by multiplying the series expansions of the exponential and of $w_i^{(\sigma)}$, Eq. (\ref{seis}). Alternatively, the same expansion results as Frobenius solution of the differential equation obeyed by $v_{i,j}^{(\sigma)}$, namely, (omitting subindices $i$ and $j$ and superindex $(\sigma)$)
\begin{eqnarray}
\fl t^2\,\frac{{\rm d}^2v(t)}{{\rm d}t^2}-2t(\beta\,t^3+\gamma\,t)\frac{{\rm d}v}{{\rm d}t}
&+&\Big((\beta\,t^3+\gamma\,t)^2-2\beta\,t^3  \nonumber   \\
& & +\,4\sigma\lambda\,t^{10}-\kappa\,t^8+4E\,t^4-3/4\Big)v(t)=0.  \label{vcinco}
\end{eqnarray}
Then, the coefficients $\hat{c}_{n,i,j}^{(\sigma)}$ can be calculated by using the recurrence relation (again omitting
subindices $i$ and $j$ and superindex $(\sigma)$)
\begin{eqnarray}
 \hat{c}_{0}&=1,\qquad \hat{c}_{2}=\gamma^2/2,  \nonumber  \\
\fl n(n-1+2\nu)\,\hat{c}_{n}&=2\gamma(n-1+\nu)\,\hat{c}_{n-1}-\gamma^2\,\hat{c}_{n-2}
+2\beta(n\! -\! 2\! +\! \nu)\,\hat{c}_{n-3} \nonumber \\
& \hspace{40pt}-\,(2\beta\gamma\! +\! 4E)\,\hat{c}_{n-4}-\beta^2\,\hat{c}_{n-6}+\kappa\,\hat{c}_{n-8}-4\sigma\lambda\,\hat{c}_{n-10}.  \label{vseis}
\end{eqnarray}
Substitution in Eq. (\ref{vuno}) of $v_{j,k}^{(\sigma)}$ by its expansion, Eq. (\ref{vcuatro}), gives
\begin{equation}
\mathcal{W}[u_{i,j}^{(\sigma)},u_j^{(\sigma)}] \sim \sum_{n=-\infty}^\infty \eta_{n,i,j}^{(\sigma)}\,t^{n+\nu_i-2},   \label{vsiete}
\end{equation}
where we have abbreviated
\begin{eqnarray}
\eta_{n,i,j}^{(\sigma)} =& \sum_{m=0}^\infty a_{m,j}^{(\sigma)}\Big(\alpha_j^{(\sigma)}\,\hat{c}_{n+m-4,i,j}^{(\sigma)}
+ 2\beta_j^{(\sigma)}\,\hat{c}_{n+m-2,i,j}^{(\sigma)}+2\gamma_j^{(\sigma)}\,\hat{c}_{n+m,i,j}^{(\sigma)} \nonumber \\ &\hspace{80pt}-\,\,(n\!+\!2m\!+\!3\!+\!\nu_i)\,\hat{c}_{n+m+1,i,j}^{(\sigma)}\Big).    \label{vocho}
\end{eqnarray}
In this way, we obtain a formal expansion, in powers of $t$, of the left-hand  side of Eq. (\ref{veinte}). Our purpose now is to obtain a similar expansion, with the same powers of $t$, for the right-hand side. Let us introduce five constants $\{g_{L,i,j}^{(\sigma)}\}$ ($L=0, 1, \ldots ,4$), to be determined, whose sum is bound by
\begin{equation}
\mathcal{W}[w_i^{(\sigma)},w_j^{(\sigma)}]=\sum_{L=0}^{4} g_{L,i,j}^{(\sigma)}\,.  \label{tuno}
\end{equation}
Then, Eq. (\ref{veinte}) becomes
\begin{equation}
\mathcal{W}[u_{i,j}^{(\sigma)},u_j^{(\sigma)}]\sim\sum_{L=0}^{4} \left(g_{L,i,j}^{(\sigma)}\,\exp\left(-\alpha_j^{(\sigma)}\,t^{5}/5\right)\right)\,.
\label{tdos}
\end{equation}
Now we replace the exponential in each one of the five terms of the right-hand side of this equation by five respective expansions
\begin{equation}
\exp\left(-\alpha_j^{(\sigma)}\,t^{5}/5\right)\sim\sum_{n=-\infty}^{\infty}\frac{\left(-\alpha_j^{(\sigma)}\,t^{5}/5\right)^{n+\delta_{L,i}}}
{\Gamma(n+1+\delta_{L,i})}, \qquad L=0, 1, \ldots , 4,   \label{vnueve}
\end{equation}
which are but particular forms of the Heaviside's exponential series \cite[Sect. 2.12]{hard},
\begin{equation}
\exp(z)\sim\sum_{n=-\infty}^{\infty}\frac{z^{n+\delta}}{\Gamma(n+1+\delta)},    \label{treinta}
\end{equation}
valid for arbitrary $\delta$, whenever $|\arg(z)|<\pi$. In view of Eqs. (\ref{vsiete}) and (\ref{vnueve}), we can write Eq. (\ref{tdos}) in the form
\begin{equation}
\sum_{m=-\infty}^\infty \eta_{m,i,j}^{(\sigma)}\,t^{m+\nu_i-2} \sim \sum_{L=0}^{4} g_{L,i,j}^{(\sigma)}\,\sum_{n=-\infty}^{\infty}\frac{\left(-\alpha_j^{(\sigma)}\,t^{5}/5\right)^{n+\delta_{L,i}}}
{\Gamma(n+1+\delta_{L,i})}\,.  \label{novisima}
\end{equation}
In order to have similar expansions in both sides of this equation, we choose for  $\delta_{L,i}$ in Eq. (\ref{vnueve}) the values
\begin{equation}
\delta_{L,i}=(\nu_i-2+L)/5, \qquad L=0, 1, \ldots, 4. \label{ttres}
\end{equation}
Then, we can compare, term by term, the resulting expansions of both sides of Eq. (\ref{novisima}) to obtain
\begin{equation}
\eta_{5n+L,i,j}^{(\sigma)} = g_{L,i,j}^{(\sigma)}\,\frac{\left(-\alpha_j^{(\sigma)}/5\right)^{n+\delta_{L,i}}}
{\Gamma(n+1+\delta_{L,i})}\,,  \qquad L=0, 1, \ldots, 4. \label{tcuatro}
\end{equation}
This equation allows one to get the values of $g_{L,i,j}^{(\sigma)}$ that, substituted in Eq. (\ref{tuno}), give
\begin{equation}
\mathcal{W}[w_{i}^{(\sigma)},w_j^{(\sigma)}]=\sum_{L=0}^{4}\frac{\Gamma(n+1+\delta_{L,i})}
{\left(-\alpha_j^{(\sigma)}/5\right)^{n+\delta_{L,i}}}\; \eta_{5n+L,i,j}^{(\sigma)}\,,
\label{tcinco}
\end{equation}
where the minus sign in front of $\alpha_j^{(\sigma)}$ is to be interpreted as $e^{i\pi}$ or $e^{-i\pi}$ so as to have
$|\arg (-\alpha_j^{(\sigma)}\,t^5)|<\pi$, which is a necessary condition to the validity of the expansions (\ref{vnueve}). Such a condition, with $\arg t=0$, cannot be fulfilled when $\alpha_j^{(\sigma)}$ is real positive; that is, in the case of positive $\lambda$, $\sigma=-1$ and $j=4$. This is not surprising, because the ray $\arg t=0$ is a Stokes ray for $T_{i,3}^{(-1)}$ when $\arg \lambda =0$. It is well known that, in the Stokes rays, the connection factors are to be taken as the average of their values in the contiguous sectors. In our case, this is equivalent to take for $\mathcal{W}[w_{i}^{(\sigma)},w_4^{(\sigma)}]$ the average of its values for $t$ slightly above and below the positive real semiaxis. The result is
\begin{equation}
\fl \mathcal{W}[w_{i}^{(-1)},w_4^{(-1)}]=(-1)^n\sum_{L=0}^{4}\cos (\delta_{L,i}\pi)\,\frac{\Gamma(n+1+\delta_{L,i})}
{\left(\alpha_4^{(-1)}/5\right)^{n+\delta_{L,i}}}\; \eta_{5n+L,i,4}^{(-1)}\,, \quad {\rm for} \;\arg \lambda=0\,.
\label{tseis}
\end{equation}

For every set of values of $\kappa$, $\lambda$ and $E$, Eqs. (\ref{docho}), (\ref{tcinco}) and (\ref{tseis}) allow one to compute the values of the Wronskians to be substituted in the expressions, Eqs. (\ref{dsiete}), giving the connection factors. From the relation
\begin{equation}
T_{1,3}^{(\sigma)}\,T_{2,4}^{(\sigma)}-T_{2,3}^{(\sigma)}\,T_{1,4}^{(\sigma)} = \frac{\mathcal{W}[w_1^{(\sigma)},w_2^{(\sigma)}]}{\mathcal{W}[w_3^{(\sigma)},w_4^{(\sigma)}]}\,,   \label{tsiete}
\end{equation}
easily derived from Eqs. (\ref{dsiete}), we can obtain directly
\begin{equation}
T_{1,3}^{(\sigma)}\,T_{2,4}^{(\sigma)}-T_{2,3}^{(\sigma)}\,T_{1,4}^{(\sigma)}=\left(\alpha_4^{(\sigma)}\right)^{-1},  \label{tocho}
\end{equation}
which provides a useful test of the accuracy in the determination of the connection factors.

\section{Case of real $\lambda$}

We mentioned at the end of section 2 that $w_3^{(+1)}(t)$ and $w_4^{(+1)}(t)$ become oscillating waves, with amplitude decreasing as $t^{-2}$, for $t\to\infty$ when $\lambda$ is real. We are then faced with a one-dimensional scattering problem. We write the solution again in the form (\ref{nueve}) with coefficients $A_1$ and $A_2$ obeying Eq. (\ref{extra1}). This condition is necessary for the cancelation of the wave function in the far left side, where the potential becomes an infinite barrier. In the far right side, instead, the solution is the superposition of an incoming (from the right) wave, $(A_1\,T_{1,4}^{(+1)}+A_2\,T_{2,4}^{(+1)})\,w_4^{(+1)}$, and an outgoing (to the right) wave, $(A_1\,T_{1,3}^{(+1)}+A_2\,T_{2,3}^{(+1)})\,w_3^{(+1)}$. If we require the values of $A_1$ and $A_2$ to fulfil both Eqs. (\ref{extra1}) and  (\ref{extra2}), or, in other words, if the energy $E$ is such that Eq. (\ref{dseis}) is satisfied, the coefficient of the incoming wave cancels. A solution with these characteristics corresponds to what is called a Gamow state. (Gamow states are frequently known as ``resonances". We prefer, however, to reserve this term to refer to solutions of the Schr\"odinger equation with real energy such that the time delay is considerably enhanced. The occurrence of a resonance is an indication of the existence of a Gamow state of complex energy, whose real part is nearly equal to the resonant energy and whose imaginary part is small. A Gamow energy of relatively large imaginary part does not imply the existence of a resonance.) The same Eq. (\ref{dseis}), giving the eigenvalues of $H$  in the case of complex $\lambda$, also gives the Gamow energies for real $\lambda$.

For any other value of $E$ we have
\begin{equation}
A_1\,T_{1,4}^{(+1)}+A_2\,T_{2,4}^{(+1)} \neq 0  , \label{tnueve}
\end{equation}
and the solution Eq. (\ref{nueve}) can be written, in the far right region, in the form
\begin{equation}
\psi_{\rm phys}(x) \sim \psi^{\rm incoming}(x) - S(E)\,\psi^{\rm outgoing}(x)\,, \qquad  {\rm for} \quad x\to +\infty\,,  \label{cuarenta}
\end{equation}
with a scattering function
\begin{equation}
S(E)= -\,\frac{T_{1,3}^{(+1)}\,T_{2,4}^{(-1)}+T_{2,3}^{(+1)}\,T_{1,4}^{(-1)}}{T_{1,4}^{(+1)}\,T_{2,4}^{(-1)}+T_{2,4}^{(+1)}\,T_{1,4}^{(-1)}}\,. \label{cuno}
\end{equation}
We remark that the scattering function presents poles at the values of the energy corresponding to Gamow states,
as it should.

\section{Eigenvalues}

We use the term ``eigenvalue" here in a rather broad sense. For complex values of $\lambda$ it refers to values of $E$ for which the Schr\"odinger equation admits solutions vanishing exponentially for $x\to\pm\infty$; the normalizable solutions are the respective eigenfunctions. For real $\lambda$, eigenvalues are the values of $E$ corresponding to Gamow states (i. e., solutions of the Schr\"odinger equation which vanish exponentially for $x\to -\infty$ and represent an outgoing wave for $x\to +\infty$).

The procedure for obtaining the eigenvalues of $H$ for given values of $\kappa$ and $\lambda$ includes the following steps:
(i) choose a value of $E$;
(ii) determine the connection factors $T_{i,j}^{(\sigma)}$ ($i=1,2$, $j=3,4$, $\sigma=+1,-1$) as indicated in section 3;
(iii) compute the left-hand  side of Eq. (\ref{dseis});
(iv) modify the value of $E$, looking for the fulfilment of Eq. (\ref{dseis}), and repeat the preceding steps until such (approximate) fulfilment is satisfactory.
This procedure has been applied to obtain the first three eigenvalues of $H$, as given in Eq. (\ref{uno}), with $\kappa=1$ and three different values of $|\lambda|$: 0.1, 1 and 10. For $\arg \lambda$ we have considered values going from 0 to $\pi/2$ by steps of $\pi/20$. The results are shown in table 1.
\begin {table}
\caption { Eigenvalues of $H(1, \lambda)$, with $|\lambda|=0.1$, $1$ and $10$, and different values of $\arg\lambda$ ranging from $0$ to $\pi/2$.}
\begin{indented}
\item[]
\hspace{-65pt}\begin{tabular}{|cc|l|l|l|}
\br
$|\lambda|$ & $(\arg \lambda)/\pi$ & 1st eigenvalue & 2nd eigenvalue & 3rd eigenvalue\\
\mr
0.1 & 0.00 & $0.415250965-0.117825303\;$i & $1.33498463-0.667795923\;$i & $2.42450721-1.40341851\;$i \\
\  & 0.05 & $0.449139561-0.112375258\;$i & $1.44066461-0.602380897\;$i & $2.59333660-1.26365922\;$i \\
\  & 0.10 & $0.477169894-0.105218927\;$i & $1.53325734-0.538939748\;$i & $2.74473328-1.12615297\;$i \\
\  & 0.15 & $0.500446382-0.096243298\;$i & $1.61304427-0.475720144\;$i & $2.87795232-0.98931955\;$i \\
\  & 0.20 & $0.519720483-0.085587341\;$i & $1.68063925-0.411647897\;$i & $2.99278233-0.85198905\;$i \\
\  & 0.25 & $0.535488315-0.073482102\;$i & $1.73670864-0.346194408\;$i & $3.08932345-0.71343381\;$i \\
\  & 0.30 & $0.548075091-0.060182762\;$i & $1.78183979-0.279212943\;$i & $3.16781588-0.57332107\;$i \\
\  & 0.35 & $0.557693501-0.045943393\;$i & $1.81649843-0.210804889\;$i & $3.22853234-0.43163385\;$i \\
\  & 0.40 & $0.564480986-0.031009625\;$i & $1.84102685-0.141225990\;$i & $3.27171631-0.28858622\;$i \\
\  & 0.45 & $0.568522698-0.015618352\;$i & $1.85565584-0.070824424\;$i & $3.29755042-0.14455243\;$i \\
\  & 0.50 & $0.569865027$  & $1.86051781$  & $3.30614914$ \\
\mr
1 & 0.00 & $0.932782218-0.667994579\;$i & $3.31926321-2.39590923\;$i & $6.11139823-4.42112766\;$i \\
\  & 0.05 & $0.975793012-0.607758070\;$i & $3.46805606-2.18191216\;$i & $6.38297781-4.02752555\;$i \\
\  & 0.10 & $1.014904552-0.545466306\;$i & $3.60314256-1.95990607\;$i & $6.62936492-3.61877643\;$i \\
\  & 0.15 & $1.049907483-0.481340615\;$i & $3.72388499-1.73074229\;$i & $6.84945501-3.19646938\;$i \\
\  & 0.20 & $1.080616412-0.415602803\;$i & $3.82971150-1.49528419\;$i & $7.04225678-2.76222533\;$i \\
\  & 0.25 & $1.106870716-0.348476231\;$i & $3.92012014-1.25440647\;$i & $7.20689918-2.31769383\;$i \\
\  & 0.30 & $1.128534882-0.280186556\;$i & $3.99468201-1.00899425\;$i & $7.34263688-1.86454946\;$i \\
\  & 0.35 & $1.145498508-0.210962082\;$i & $4.05304368-0.75994203\;$i & $7.44885450-1.40448812\;$i \\
\  & 0.40 & $1.157676124-0.141033744\;$i & $4.09492881-0.50815229\;$i & $7.52507038-0.93922332\;$i \\
\  & 0.45 & $1.165006932-0.070634755\;$i & $4.12013931-0.25453388\;$i & $7.57093867-0.47048197\;$i \\
\  & 0.50 & $1.167454568$ & $4.12855606$ & $7.58625127$ \\
\mr
10 & 0.00 & $2.34949018-1.70643268\;$i & $8.35021806-6.06584226\;$i & $15.3672631-11.1638105\;$i \\
\  & 0.05 & $2.54218953-1.55549565\;$i & $8.71493106-5.52948200\;$i & $16.0383050-10.1767683\;$i \\
\  & 0.10 & $2.54521325-1.39844366\;$i & $9.04525428-4.97133937\;$i & $16.6460562-9.14961233\;$i \\
\  & 0.15 & $2.62818978-1.23589638\;$i & $9.33987559-4.39361705\;$i & $17.1881090-8.08639607\;$i \\
\  & 0.20 & $2.70078734-1.06849435\;$i & $9.59762793-3.79859361\;$i & $17.6623152-6.99131395\;$i \\
\  & 0.25 & $2.76271542-0.89689653\;$i & $9.81748419-3.18861460\;$i & $18.0667946-5.86868436\;$i \\
\  & 0.30 & $2.81372597-0.72177775\;$i & $9.99857144-2.56608338\;$i & $18.3999435-4.72293312\;$i \\
\  & 0.35 & $2.85361457-0.54382613\;$i & $10.1401696-1.93345179\;$i & $18.6604401-3.55857563\;$i \\
\  & 0.40 & $2.88222128-0.36374050\;$i & $10.2417155-1.29321065\;$i & $18.8472511-2.38020002\;$i \\
\  & 0.45 & $2.89943135-0.18222771\;$i & $10.3028051-0.64788013\;$i & $18.9596349-1.19244856\;$i \\
\  & 0.50 & $2.90517572$ & $10.3231953$ & $18.9971458$ \\
\br
\end{tabular}
\end{indented}
\end{table}

Very precise eigenvalues, obtained by using complex scaling and diagonalization of the complex-scaled operator in the basis of harmonic oscillator wavefunctions, were published by Alvarez \cite[table I]{alva1}, for the Hamiltonian
\begin{equation}
H_A=\frac{1}{2}\,p^2+\frac{1}{2}\,k \,x^2+g\,x^3,   \label{alvarez}
\end{equation}
with $k=1/4$ and different values of $g$. For comparison, we have taken in Eq. (\ref{uno}) $\kappa=1$ and $\lambda=2g$. Then, our eigenenergies should be double those reported in Ref. \cite{alva1}. Our results agree with those of Alvarez, with the exception of one entry, were a misprint (we suppose) has transposed two contiguous digits. (For $g=0.5$, the reported value of $\Re E$ is 0.4663911098243, whereas, according to our results, it should be 0.4663911089243.)

We have already mentioned in section 1  that the eigenvalues of $H(\kappa, \lambda)$ with a nonzero value of $\kappa$ different from $1$ are obtained from those for $\kappa=1$ with an adequate scaling of $|\lambda|$.
Obviously, the eigenvalues of $H$ with $\kappa=0$ cannot be obtained by a mere scaling of the problem with $\kappa=1$. It is, then, unavoidable that one must directly solve the differential equation (\ref{dos}) with $\kappa=0$ and an arbitrarily selected value of $\lambda$. By choosing $\lambda={\rm i}$, an infinite set, $\{E_n(\textrm{i})\}$, of real eigenvalues is obtained \cite{bedo}. The lowest of them are
\[
\fl E_1(\lambda=\textrm{i})=1.156267072\,,\quad E_2(\lambda=\textrm{i})=4.109228753\,, \quad     E_3(\lambda=\textrm{i})=7.56227361\,.
\]
For any other value of $\lambda$, complex scaling of the variable $x$ (Symanzik scaling) makes evident that the eigenvalues $E_n(\lambda)$
of $H(0, \lambda)$ and those of $H(0, \textrm{i})$ are related by
\begin{equation}
E_n(\lambda) = \left(\lambda\,\textrm{e}^{-{\rm i}\pi/2}\right)^{2/5}\,E_n(\textrm{i})\,  \label{v1}
\end{equation}
and consequently
\[
\left|E_n(\lambda)\right|=|\lambda|^{2/5}\,E_n(\textrm{i})\,,  \qquad \arg E_n(\lambda)=\frac{2}{5}\left(\arg \lambda-\frac{\pi}{2}\right)\,.
\]
We show, in table 2, the eigenvalues of $H$ when $\kappa=0$, for the values of $\lambda$ considered in table 1. It can be seen that the effect of the quadratic term is very small, and the eigenvalues are determined almost entirely by the cubic term.
\begin {table}
\caption { Eigenvalues of $H(0, \lambda)$, with $|\lambda|=0.1$, $1$ and $10$, and different values of $\arg\lambda$ ranging from $0$ to $\pi/2$.}
\begin{indented}
\item[]
\hspace{-65pt}\begin{tabular}{|cc|l|l|l|}
\br
$|\lambda|$ & $(\arg \lambda)/\pi$ & 1st eigenvalue & 2nd eigenvalue & 3rd eigenvalue\\
\mr
0.1 &0.00 & $0.372405257-0.270568257\;$i & $1.32348177-0.961565789\;$i & $2.43562280-1.76958355\;$i \\
\  & 0.05 & $0.388659521-0.246650832\;$i & $1.38124739-0.876566251\;$i & $2.54192973-1.61315766\;$i \\
\  & 0.10 & $0.403379925-0.221759990\;$i & $1.43356187-0.788107306\;$i & $2.63820482-1.45036537\;$i \\
\  & 0.15 & $0.416508371-0.195993962\;$i & $1.48021873-0.696538062\;$i & $2.72406812-1.28184916\;$i \\
\  & 0.20 & $0.427993050-0.169454436\;$i & $1.52103384-0.602219900\;$i & $2.79918076-1.10827407\;$i \\
\  & 0.25 & $0.437788636-0.142246150\;$i & $1.55584613-0.505525052\;$i & $2.86324633-0.93032513\;$i \\
\  & 0.30 & $0.445856469-0.114476484\;$i & $1.58451820-0.406835127\;$i & $2.91601196-0.74870462\;$i \\
\  & 0.35 & $0.452164711-0.086255032\;$i & $1.60693691-0.306539609\;$i & $2.95726943-0.56412931\;$i \\
\  & 0.40 & $0.456688466-0.057693170\;$i & $1.62301377-0.205034320\;$i & $2.98685591-0.37732765\;$i \\
\  & 0.45 & $0.459409880-0.028903620\;$i & $1.63268533-0.102719854\;$i & $3.00465463-0.18903685\;$i \\
\  & 0.50 & $0.460318212$ & $1.63591343$ & $3.01059535$ \\
\mr
1 & 0.00 & $0.935439711-0.679636733\;$i & $3.32443590-2.41534406\;$i & $6.11800786-4.44499290\;$i \\
\  & 0.05 & $0.976268578-0.619558879\;$i & $3.46953659-2.20183487\;$i & $6.38503879-4.05206883\;$i \\
\  & 0.10 & $1.013244559-0.557035910\;$i & $3.60094461-1.97963605\;$i & $6.62687088-3.64315310\;$i \\
\  & 0.15 & $1.046221727-0.492314575\;$i & $3.71814134-1.74962451\;$i & $6.84254974-3.21985950\;$i \\
\  & 0.20 & $1.075069935-0.425650299\;$i & $3.82066427-1.51270800\;$i & $7.03122418-2.78385859\;$i \\
\  & 0.25 & $1.099675333-0.357406175\;$i & $3.90810878-1.26982152\;$i & $7.19214959-2.33687106\;$i \\
\  & 0.30 & $1.119940816-0.287551928\;$i & $3.98012978-1.02192364\;$i & $7.32469088-1.88066097\;$i \\
\  & 0.35 & $1.135786403-0.216662844\;$i & $4.03644314-0.76999269\;$i & $7.42832495-1.41702877\;$i \\
\  & 0.40 & $1.147149561-0.144918691\;$i & $4.07682626-0.51502293\;$i & $7.50264282-0.94780422\;$i \\
\  & 0.45 & $1.153985443-0.072602610\;$i & $4.10112013-0.25802061\;$i & $7.54735119-0.47483909\;$i \\
\  & 0.50 & $1.156267072$ & $4.10922875$ & $7.56227361$ \\
\mr
10 & 0.00 & $2.34971832-1.70717029\;$i & $8.35060542-6.06706997\;$i & $15.3677410-11.1653174\;$i \\
\  & 0.05 & $2.45227580-1.55626154\;$i & $8.71508188-5.53075914\;$i & $16.0384923-10.1783367\;$i \\
\  & 0.10 & $2.54515526-1.39921094\;$i & $9.04516390-4.97262093\;$i & $16.6459470-9.15118683\;$i \\
\  & 0.15 & $2.62799016-1.23663830\;$i & $9.33954878-4.39485806\;$i & $17.1877079-8.08792139\;$i \\
\  & 0.20 & $2.70045358-1.06918521\;$i & $9.59707438-3.79975069\;$i & $17.6616366-6.99273662\;$i \\
\  & 0.25 & $2.76225955-0.89751253\;$i & $9.81672542-3.18964744\;$i & $18.0658630-5.86995471\;$i \\
\  & 0.30 & $2.81316414-0.72229779\;$i & $9.99763398-2.56695611\;$i & $18.3987916-4.72400677\;$i \\
\  & 0.35 & $2.85296646-0.54423246\;$i & $10.1390864-1.93413418\;$i & $18.6591087-3.55941534\;$i \\
\  & 0.40 & $2.88150942-0.36401929\;$i & $10.2405246-1.29367910\;$i & $18.8457867-2.38077652\;$i \\
\  & 0.45 & $2.89868038-0.18236951\;$i & $10.3015480-0.64811846\;$i & $18.9580891-1.19274186\;$i \\
\  & 0.50 & $2.90441157$ & $10.3219160$ & $18.9955727$ \\
\br
\end{tabular}
\end{indented}
\end{table}

Our results confirm the analyticity of the eigenvalues, considered as a function of $\lambda$, discussed by Zinn-Justin and Jentschura \cite{zinn} and by Grecchi  and Martinez \cite{grec}. In the case of a pure cubic potential ($\kappa=0$), Eq. (46) shows that the only singularity is a branch point at $\lambda=0$. When an $x^2$ term is added to the Hamiltonian ($\kappa\neq 0$), the analyticity is not destroyed. In fact, the eigenvalues are determined by the behavior of the wave function at the boundaries, just the regions where the cubic term dominates. So, a perturbative treatment of the problem would require one to consider the harmonic term as a perturbation to the pure cubic Hamiltonian. Unfortunately, we do not dispose of a complete set of eigenfunctions of this Hamiltonian, as we do for the harmonic oscillator, and all sensible perturbative calculations have taken the harmonic oscillator Hamiltonian as the unperturbed one and the cubic term as a perturbation. The price to be paid is the divergence of the conventional perturbative series.

\section{Eigenfunctions}

Once the eigenvalues have been obtained, it becomes trivial to get the corresponding wave functions.
The coefficients $A_1$ and $A_2$ in Eq. (\ref{nueve}) are determined, up to a common arbitrary multiplicative constant,
by solving either Eq. (\ref{extra1}) or Eq. (\ref{extra2}). Let us take the second one. To fix the arbitrary constant
we require, for convenience, that
\begin{equation}
A_1\,T_{1,3}^{(+1)}+A_2\,T_{2,3}^{(+1)} = \textrm{e}^{{\rm i}\,3\pi/8}\,.   \label{cuarenta}
\end{equation}
Then, we obtain
\begin{equation}
A_1=\frac{\textrm{e}^{{\rm i}\,3\pi/8}\,T_{2,4}^{(+1)}}{T_{1,3}^{(+1)}T_{2,4}^{(+1)}-T_{2,3}^{(+1)}T_{1,4}^{(+1)}}, \qquad
A_2=\frac{-\,\textrm{e}^{{\rm i}\,3\pi/8}\,T_{1,4}^{(+1)}}{T_{1,3}^{(+1)}T_{2,4}^{(+1)}-T_{2,3}^{(+1)}T_{1,4}^{(+1)}},   \label{cuno}
\end{equation}
or, in view of Eq. (\ref{tocho}),
\begin{equation}
A_1=\textrm{e}^{{\rm i}\,3\pi/8}\,\alpha_4^{(+1)}\,T_{2,4}^{(+1)}\,, \qquad
A_2=-\,\textrm{e}^{{\rm i}\,3\pi/8}\,\alpha_4^{(+1)}\,T_{1,4}^{(+1)} \,.   \label{extra3}
\end{equation}
Then, $w_{\rm phys}(t)$ is obtained as a linear combination of the series given in Eq. (\ref{seis}). Finally, Eqs. (\ref{tres}) and (\ref{cuatro}) allow one to write the wave function
\begin{equation}
\psi_{\rm phys}(x)=\mathcal{N}\,\sum_{n=0}^\infty\,d_n\,x^n,  \label{ccuatro}
\end{equation}
where $\mathcal{N}$ represents a normalization constant such that
\begin{equation}
\int_{-\infty}^\infty|\psi_{\rm phys}(x)|^2\,dx = 1\,,  \label{quno}
\end{equation}
 and the coefficients $d_n$ are given by the recurrence relation
\begin{equation}
\fl d_0=A_1,\qquad d_1=A_2, \qquad
 2n(2n-2)\,d_n=-\,4E\,d_{n-2}+\kappa\,d_{n-4}-4\lambda\,d_{n-5}. \label{ccinco}
\end{equation}
Although the series in the right-hand  side of Eq. (\ref{ccuatro}) is convergent in all the finite $x$-plane, this expression of $\psi_{\rm phys}(x)$ is not useful, from the computational point of view, for large positive or negative values of $x$. It becomes more convenient, in these cases, to use the asymptotic expansions stemming from the Thom\'e solutions, as seen in Eqs. (\ref{diez}). In fact, for large positive $x$, bearing in mind Eq. (\ref{cuarenta}),
\begin{equation}
\fl \psi_{\rm phys}(x) \sim \mathcal{N}\,\textrm{e}^{{\rm i}\,3\pi/8}\,\exp\left(B^{(+)}(x)\right)\,x^{-3/4}
\,\sum_{m=0}^\infty a_{m,3}^{(+1)}\,x^{-m/2}, \qquad x\to +\infty\,,  \label{cseis}
\end{equation}
where we have abbreviated
\begin{equation}
B^{(+)}(x) \equiv {\rm i}\,\left(\frac{2}{5}\,\lambda^{1/2}x^{5/2}-\frac{\kappa}{12}\,\lambda^{-1/2}x^{3/2}-\frac{\kappa^2}{64}\,\lambda^{-3/2}x^{1/2}\right) ,
\label{csiete}
\end{equation}
and the coefficients  $a_{m,3}^{(+1)}$ are given by Eq. (\ref{ddos}) with $\alpha$ replaced by $\alpha_3^{(+1)}$.
For large negative values of $x$ we need  to calculate the value of $A_1\,T_{1,3}^{(-1)}-A_2\,T_{2,3}^{(-1)}$, which, in view of Eqs. (\ref{dseis}) and (\ref{tocho}), satisfies
\begin{equation}
A_1\,T_{1,3}^{(-1)}-A_2\,T_{2,3}^{(-1)}=\textrm{e}^{{\rm i}\,3\pi/8}\,\frac{\alpha_4^{(+1)}\,T_{2,4}^{(+1)}}{\alpha_4^{(-1)}\,T_{2,4}^{(-1)}} \label{cocho}
\end{equation}
and turns out to be
\begin{equation}
A_1\,T_{1,3}^{(-1)}-A_2\,T_{2,3}^{(-1)}=\textrm{e}^{-{\rm i}\,3\pi/8}.   \label{cnueve}
\end{equation}
Then in the far left region the wave function can be calculated by using its asymptotic expression
\begin{equation}
\fl \psi_{\rm phys}(x) \sim \mathcal{N}\,\textrm{e}^{-{\rm i}\,3\pi/8}\,\exp\left(B^{(-)}(x)\right)\,(-x)^{-3/4}
\,\sum_{m=0}^\infty a_{m,3}^{(-1)}\,(-x)^{-m/2}, \qquad x\to -\infty\,,  \label{cincuenta}
\end{equation}
with
\begin{equation}
B^{(-)}(x) \equiv -\,\left(\frac{2}{5}\,\lambda^{1/2}(-x)^{5/2}+\frac{\kappa}{12}\,\lambda^{-1/2}(-x)^{3/2}-\frac{\kappa^2}{64}\,\lambda^{-3/2}(-x)^{1/2}\right).
\label{nueva}
\end{equation}

We show, in figure 1, the squared modulus of the first eigenfunction of the Hamiltonian given by Eq. (\ref{uno}), with $\kappa=1$, $|\lambda|=1$ and three different values of $\arg \lambda$. The eigenfunctions have been normalized according to Eq. (\ref{quno}). The corresponding eigenvalues and the parameters needed to evaluate the eigenfunctions are given in table 3. Notice that, contrary to what happens for short-range potentials, the Gamow state wave functions (in the case of $\arg\lambda=0$) are normalizable. Notice, also, that in a time-dependent formulation of the problem, the probability densities shown in figure 1 keep their shape but vanish exponentially as time goes on, except in the case of $\arg\lambda=\pi/2$ (real eigenenergy). In this case, thanks to our choice of the right-hand  side of Eq. (\ref{cuarenta}), the arbitrary phase of the eigenfunction has been fixed in such a way that
\begin{equation}
\psi_{\rm phys}(-x)=(\psi_{\rm phys}(x))^*,  \label{qdos}
\end{equation}
where the asterisk indicates complex conjugation.
\begin{table}
\caption{Eigenenergy, parameters of the wave function and normalization constant of the first eigenstate of the Hamiltonian Eq. (\ref{uno}), with $\kappa=1$, $|\lambda|=1$, and three different values of $\arg \lambda$.}
\begin{indented}
\item[]
\begin{tabular}{lcrl}
\br
$\arg\lambda=0$ & \ & \ & \  \\
\ & \ & $E=$ & 0.9327822178 $-$ 0.6679945787 i \\
\  & \ & \ & \  \\
\  & \ &  $A_1=$ & 0.4628577159 $-$ 0.1381312181 i \\
\  & \ &  $A_2=$ & 0.1664674468 $+$ 0.1856562056 i \\
\  & \ & \ & \  \\
\  & \ &  $\mathcal{N}=$ &  0.81730935  \\
\mr
$\arg\lambda=\pi/4$ & \ & \ & \  \\
\ & \ & $E=$ & 1.106870716 $-$ 0.3484762310 i \\
\  & \ & \ & \  \\
\  & \ &  $A_1=$ & 0.4551986334 $-$ 0.0755253335 i \\
\  & \ &  $A_2=$ & 0.0866424469 $+$ 0.2056069342 i \\
\  & \ & \ & \  \\
\  & \ &  $\mathcal{N}=$ &  1.55999007  \\
\mr
$\arg\lambda=\pi/2$ & \ & \ & \  \\
\ & \ & $E=$ & 1.167454568 \\
\  & \ & \ & \  \\
\  & \ &  $A_1=$ & 0.4506119146 \\
\  & \ &  $A_2=$ & 0.2126331531 i \\
\  & \ & \ & \  \\
\  & \ &  $\mathcal{N}=$ &  1.66344664  \\
\br
\end{tabular}
\end{indented}
\end{table}
\begin{figure}
\begin{center}
\includegraphics{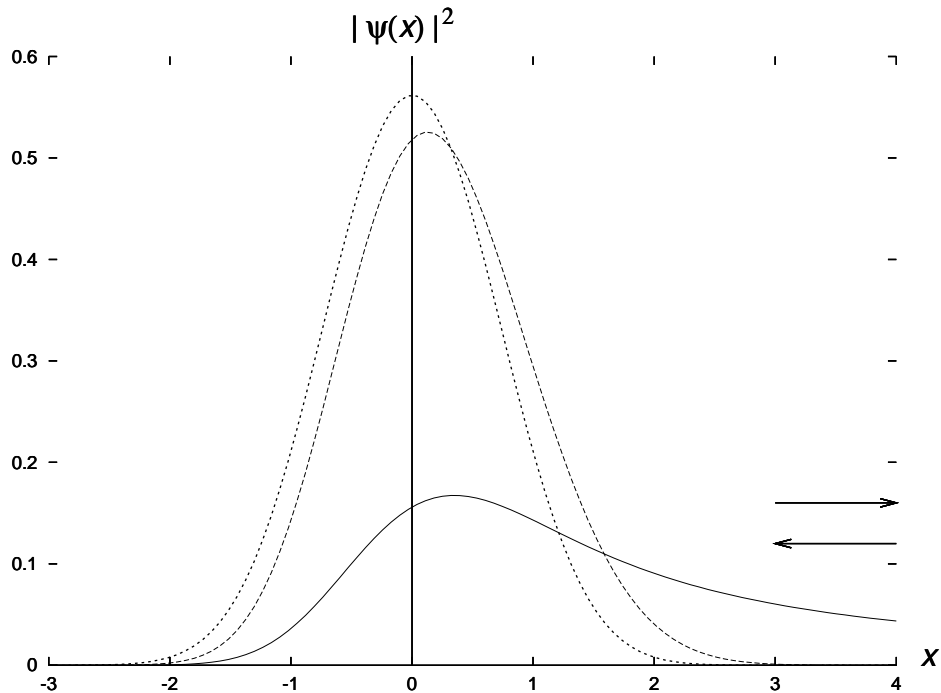}
\end{center}
\caption{Squared modulus of the normalized first eigenfunction of the Hamiltonian $H(1, \lambda)$, with $|\lambda|=1$ and three different values of $\arg\lambda$: $0$ (continuous line), $\pi/4$ (dashed line), and $\pi/2$ (dotted line). For $x\to \pm\infty$, $|\psi(x)|^2$ vanishes exponentially for every value of $\arg\lambda$, with the exception of $\arg\lambda=0$, in which case $|\psi(x)|^2$ goes to zero like $x^{-3/2}$ as $x\to+\infty$. The horizontal arrows indicate the directions of the incoming and outgoing waves in the case of real $\lambda$} \label{cub1}
\end{figure}

Some comments about our interpretation of the squared modulus of the wave function are in order. The use of complex potentials has a long tradition in nuclear physics as a way of describing, in the entrance channel, the effects of the open reaction channels. Here we adhere to the widely accepted interpretation of $|\psi(x)|^2$ as a density probability, which is consistent with its necessary positivity and with the common formulation of the continuity equation. Notice, however, that in other approaches to the cubic oscillator problem such as that of Jentschura {\em et al} \cite{jent}, which look for a complete set of eigenfunctions by diagonalization of the matrix representing the complex scaled Hamiltonian in a harmonic oscillator basis, the scalar product $({\bf \Psi}_n|{\bf \Psi}_m)$ of two of those eigenfunctions, represented by column matrices ${\bf \Psi}_n$ and ${\bf \Psi}_m$, is obtained by summing the products of homologous elements of the column matrices, without need of complex conjugation of the elements of the first column. The issue, raised by Moiseyev and collaborators \cite{mois}, has been addressed recently by Noble, Lubasch and Jentschura \cite{nobl}, who have also given a very efficient algorithm for the diagonalization of complex symmetric matrices.

We have seen, in the preceding section, that the eigenvalues of the cubic oscillator are determined mainly by the cubic term. An analogous conclusion about the eigenfunctions is apparent from a comparison of the graphics in figure 1 and the parameters in table 3 with those corresponding to a pure cubic potential ($\kappa=0$) with the same values of $\lambda$, given in figure 2 and table 4.
\begin{table}
\caption{Eigenenergy, parameters of the wave function and normalization constant of the first eigenstate of the Hamiltonian Eq. (\ref{uno}), with $\kappa=0$, $|\lambda|=1$, and three different values of $\arg \lambda$.}
\begin{indented}
\item[]
\begin{tabular}{lcrl}
\br
$\arg\lambda=0$ & \ & \ & \  \\
\ & \ & $E=$ & 0.9354397113 $-$ 0.6796367326 i \\
\  & \ & \ & \  \\
\  & \ &  $A_1=$ & 0.4681342359 $-$ 0.1123890864 i \\
\  & \ &  $A_2=$ & 0.1355137918 $+$ 0.2211384098 i \\
\  & \ & \ & \  \\
\  & \ &  $\mathcal{N}=$ &  0.82440083  \\
\mr
$\arg\lambda=\pi/4$ & \ & \ & \  \\
\ & \ & $E=$ & 1.099675333 $-$ 0.3573061753 i \\
\  & \ & \ & \  \\
\  & \ &  $A_1=$ & 0.4780992640 $-$ 0.0565867769 i \\
\  & \ &  $A_2=$ & 0.0704000491 $+$ 0.2496197451 i \\
\  & \ & \ & \  \\
\  & \ &  $\mathcal{N}=$ &  1.47311606  \\
\mr
$\arg\lambda=\pi/2$ & \ & \ & \  \\
\ & \ & $E=$ & 1.156267072 \\
\  & \ & \ & \  \\
\  & \ &  $A_1=$ & 0.4814363609 \\
\  & \ &  $A_2=$ & 0.2593572517 i \\
\  & \ & \ & \  \\
\  & \ &  $\mathcal{N}=$ &  1.52834769  \\
\br
\end{tabular}
\end{indented}
\end{table}
\begin{figure}
\begin{center}
\includegraphics{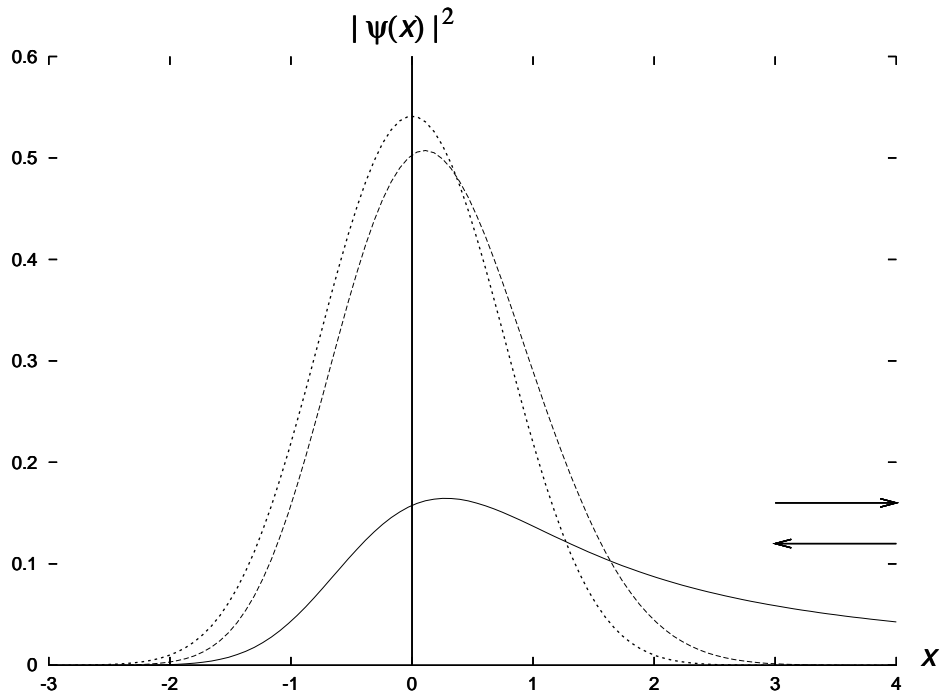}
\end{center}
\caption{Squared modulus of the normalized first eigenfunction of the Hamiltonian $H(0, \lambda)$, with $|\lambda|=1$ and three different values of $\arg\lambda$: $0$ (continuous line), $\pi/4$ (dashed line), and $\pi/2$ (dotted line). The behavior of $|\psi(x)|^2$ for $x\to\pm\infty$ is analogous to that in figure 1.}
\label{cub2}
\end{figure}

\section{Scattering by a cubic real potential}

In the two preceding sections we have shown the procedure for finding the eigenvalues and eigenfunctions of the  Hamiltonian Eq. (\ref{uno}) for arbitrary values of $\lambda$ and, in particular, for real $\lambda$. Now the object of this section is to discuss the scattering problem
for the case of real $\lambda$. It is important to stress that scattering can occur only if $\lambda$ is real. An imaginary part in this parameter implies an exponential vanishing of the wave functions at large distances, as shown in Eqs. (\ref{cseis}) and (\ref{cincuenta}), and therefore, the existence of bound states instead of scattering ones.

We have already given, in Eq. (\ref{cuno}), the scattering function in the form
\begin{equation}
S(E)=-\, \frac{N(E)}{D(E)}\,,  \label{qtres}
\end{equation}
with
\begin{equation}
\fl N(E)=T_{1,3}^{(+1)}\,T_{2,4}^{(-1)}+T_{2,3}^{(+1)}\,T_{1,4}^{(-1)}\,,\qquad D(E)=T_{1,4}^{(+1)}\,T_{2,4}^{(-1)}+T_{2,4}^{(+1)}\,T_{1,4}^{(-1)}\,. \label{qcuatro}
\end{equation}
From the structure of the connection factors one can see that, for real $\lambda$,
\begin{equation}
N(E^*)=[D(E)]^*\,, \qquad {\rm and} \qquad  D(E^*)=[N(E)]^*\,,   \label{qcinco}
\end{equation}
which imply the fulfilment of the familiar unitarity condition
\begin{equation}
S(E)[S(E^*)]^*=1\,.   \label{qseis}
\end{equation}

For real values of the energy, $E$, it is possible to define a (real) phase shift, $\delta(E)$, such that
\begin{equation}
S(E)=\exp[2{\rm i}\,\delta(E)].   \label{qsiete}
\end{equation}
As in the case of short-range potentials, this definition suffers from ambiguity: the value of $\delta(E)$ is determined up to addition of $n\pi$ ($n$ integer).
The time delay suffered by the incident wave in its interaction with the potential is closely related to the phase shift \cite[pp 110--111]{nuss},
\begin{equation}
\Delta\tau=2 \hbar\,\frac{{\rm d}\delta (E)}{{\rm d}E}\,.  \label{qocho}
\end{equation}
In Figs. 3 and 4, we show these two quantities for a particular potential related to one thoroughly discussed by Fern\'andez and Garc\'{\i}a \cite{fega},
 who  have considered, among others, the Hamiltonian
\begin{equation}
H_3=\frac{1}{2}\,p^2-x^3\,,    \label{qnueve}
\end{equation}
determining the six lowest eigenvalues  with high  accuracy. In order to compare results, we have taken, in the Hamiltonian of Eq. (\ref{uno}),
the parameter values $\kappa =0$ and $\lambda =2$. Then, the eigenvalues of $H$ are twice those of $H_3$. Accordingly, the energy of the first Gamow state of $H$ turns out to be (keeping only ten digits of the value given in Ref. \cite{fega}) $1.2343200991-0.8967860451\,$i. The existence of a resonance for $E\approx 1.26$ is revealed, in figure 4, by an enhancement of the time delay in the neighbourhood of this value. Apparently, there is a second hump that one would be tempted to associate with the second Gamow state, of energy $4.386619462-3.186065593\,$i. However, a careful examination of the data shows that the value of $\Delta\tau$ is monotonously decreasing and the hump does not exist. Other Gamow states do not seem to have any influence on the time delay.
\begin{figure}
\vspace{1cm}
\begin{center}
\includegraphics{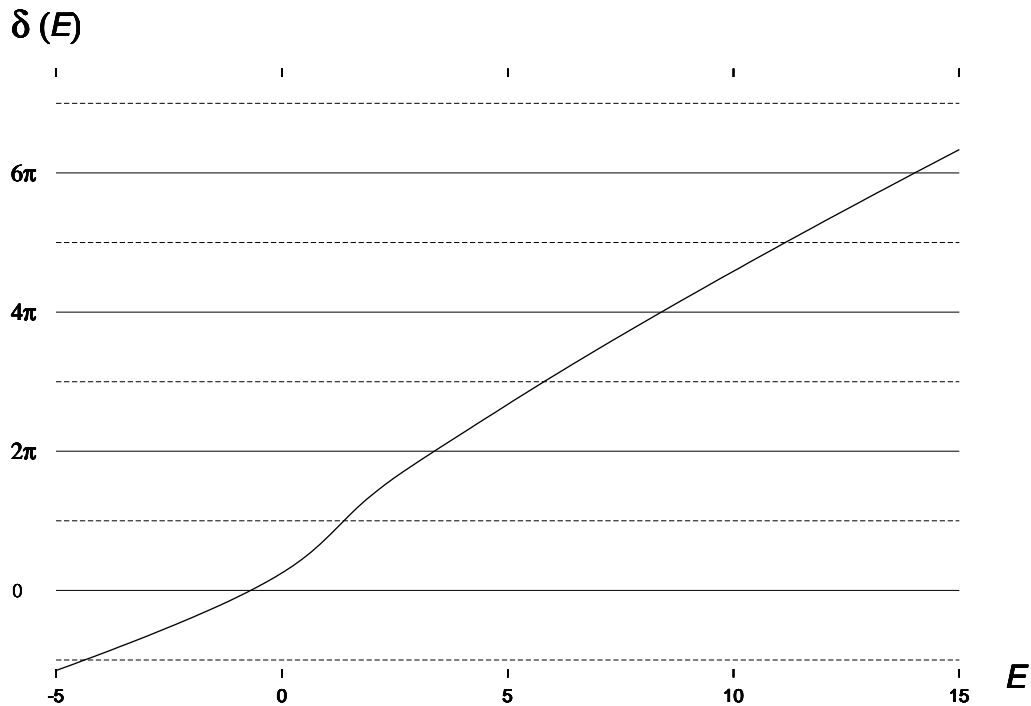}
\end{center}
\caption{Phase shift $\delta(E)$ of a particle of energy $E$ incident from the right and scattered back by the potential $-2 x^3$. To eliminate the ambiguity in the definition of $\delta$, we have chosen the interval $[0, \pi)$ to contain the value of $\delta(0)$.} \label{cub3}
\end{figure}
\begin{figure}
\vspace{1cm}
\begin{center}
\includegraphics{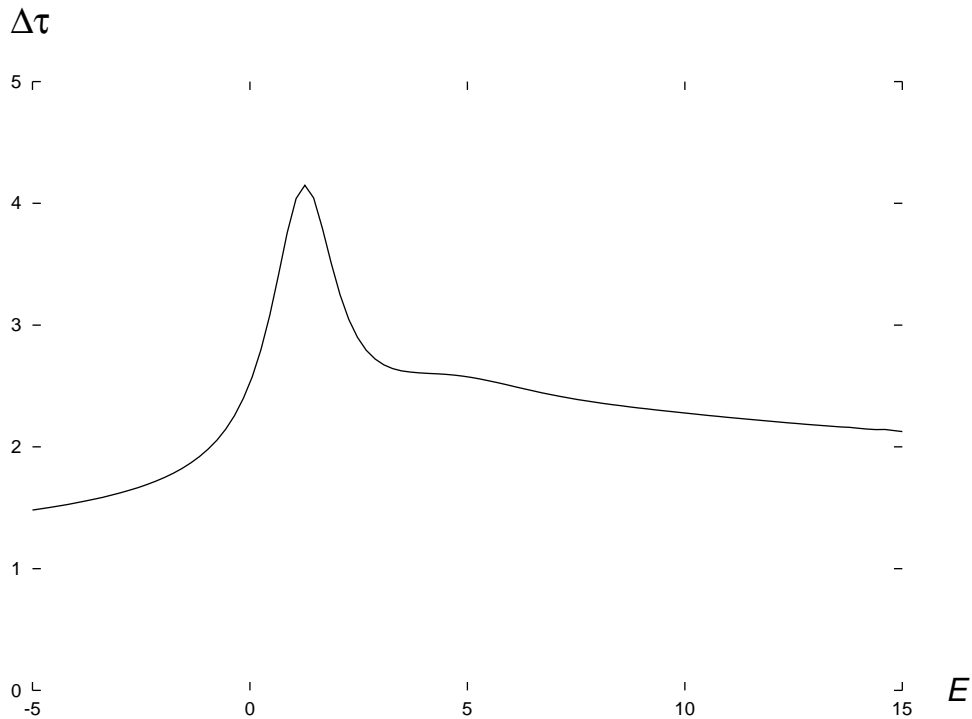}
\end{center}
\caption{Time delay suffered by a particle of energy $E$  scattered by the potential $-2 x^3$. The scale is consistent with that used for lengths and energies.} \label{cub4}
\end{figure}

\section{Final comments}

We have presented a new procedure to deal with the one-dimensional cubic anharmonic  oscillator. The Schr\"odinger equation is  solved directly,
without use of perturbation expansions, variational techniques, or diagonalization in harmonic oscillator bases.
Besides the determination of  eigenvalues, the procedure gives the eigenfunctions as power series of $x$ that are convergent for all finite values of the variable, although they should be used to compute the eigenfunctions only for small and moderate values of $|x|$. In the regions of large $|x|$,
 asymptotic expansions, provided   by the same procedure, are more convenient from the computational point of view.

In the case of complex values of the coefficient $\lambda$ of the cubic term, the eigenstates become confined: their probability density vanishes exponentially as $x\to\pm\infty$. This confinement, an effect that has been studied for many years \cite{gom3}, may be understood as due to the absorptive nature of the complex potential. For real $\lambda$, the eigenstates are Gamow states. The fact that the potential has infinite range makes these states to have properties different from those found in the case of short-range potentials. This circumstance was already discussed in a former paper \cite{ferr}, where the scattering by the potential
\begin{equation}
V(x)=\left\{\begin{array}{lll}x^2,&\qquad \mbox{for}&\quad x<0, \\ -\,x^2, &\qquad \mbox{for}&\quad x>0,\end{array}\right.  \label{i4}
\end{equation}
was considered. It was shown that the amplitude of the outgoing wave of the Gamow states behaves, for $x\to\infty$, as $x^{-(1-{\rm i}E_G)/2}$,
$E_G$ being the energy of the state, instead of increasing exponentially, as it happens in the case of short-range potentials \cite{garc}. Here, for the Hamiltonian Eq. (\ref{uno}), we have found that the amplitude of the Gamow outgoing waves decreases as $x^{-3/4}$, independently of the value of the energy, and the Gamow states, although not confined, become normalizable. This last property is preserved for steeper potentials like, for instance, those considered in \cite{fega},
\begin{equation}
V_K(x)=-\,x^K, \qquad K=5, 7, \ldots\,.  \label{i5}
\end{equation}
The amplitude of the Gamow outgoing waves decreases in these cases as $x^{-K/4}$. The physical reasons of such damping of the probability density
as $x\to\infty$ have also been discussed   \cite{ferr}.

\ack{The idea of applying the modified Naundorf's method for solving the cubic oscillator problem was suggested by Prof. Alberto Galindo. Financial support from Conselho Nacional de Desenvolvimento Cient\'{\i}fico e Tecnol\'{o}gico  (CNPq, Brazil) and from Departamento de Ciencia, Tecnolog\'{\i}a y Universidad del Gobierno de Arag\'on (Project E24/1) and Ministerio de Ciencia e Innovaci\'on (Project MTM2012-33575) is gratefully acknowledged.}


\section*{References}

\begin{thebibliography}{99}

\bibitem{davi} Davydov A S 1965 {\it Quantum Mechanics} (Oxford: Pergamon Press)

\bibitem{yari} Yaris R, Bendler J, Lovett R A, Bender C M and Fedders P A 1978 {\it Phys. Rev. A} {\bf 18} 1816

\bibitem{cali} Caliceti E, Graffi S and Maioli M 1980 {\it Commun. Math. Phys.} {\bf 75} 51

\bibitem{alva1} Alvarez G 1988 {\it Phys. Rev. A} {\bf 37} 4079

\bibitem{alva2} Alvarez G 1989 {\it J. Phys. A: Math. Gen.} {\bf 22} 617

\bibitem{alva3} Alvarez G 1995 {\it J. Phys. A: Math. Gen.} {\bf 27} 4589

\bibitem{alva4} Alvarez G and Casares C 2000 {\it J. Phys. A: Math. Gen.} {\bf 33} 2499

\bibitem{alva5} Alvarez G and Casares C 2000 {\it J. Phys. A: Math. Gen.} {\bf 33} 5171

\bibitem{jent} Jentschura U D, Surzhykov A, Lubasch M and Zinn-Justin J 2008 {\it J. Phys. A: Math. Theor.} {\bf 41} 095302

\bibitem{bend1} Bender C M 2005 {\it Contemp. Phys.} {\bf 46} 277
    \nonum Bender C M 2007 {\it Rep. Progr. Phys.} {\bf 70} 947

\bibitem{fern} Fern\'andez F M, Guardiola R, Ros J and Znojil M 1998 {\it J. Phys. A: Math. Gen.} {\bf 31} 10105
    \nonum Bender C M and Dunne G V 1999 {\it J. Math. Phys.} {\bf 40} 4616
    \nonum Grecchi V, Maioli M and Martinez A 2009 {\it J. Phys. A: Math. Theor.} {\bf 42} 425208
    \nonum Grecchi V, Maioli M and Martinez A 2010 {\it J. Phys. A: Math. Theor.} {\bf 43} 474027

\bibitem{bend2} Bender C M and Mannheim P D 2010 {\it Phys. Lett. A} {\bf 374} 1616

\bibitem{voro} Voros A 1999 {\it J. Phs. A: Math. Gen.} {\bf 32} 1301

\bibitem{zin1} Zinn-Justin J and Jentschura U D 2004 {\it Ann. Phys., NY} {\bf 313} 197
     \nonum Zinn-Justin J and Jentschura U D 2004 {\it Ann. Phys., NY} {\bf 313} 269
     \nonum Jentschura U D, Surzhykov A and Zinn-Justin J 2010 {\it Ann. Phys., NY} {\bf 325} 1135

\bibitem{prl} Jentschura U D, Surzhykov A, Lubasch M and Zinn-Justin J 2009 {\it Phys. Rev. Lett.} {\bf 102} 011601

\bibitem{zinn} Zinn-Justin J and Jentschura U D 2010 {\it J. Phys. A: Math. Theor.} {\bf 43} 425301

\bibitem{grec} Grecchi V and Martinez A 2013 {\it Commun. Math. Phys.} {\bf 319} 479

\bibitem{fega} Fern\'andez F M and Garc\'{\i}a J 2013 {\it J. Phys. A: Math. Theor.} {\bf 46} 195301

\bibitem{naun} Naundorf F 1974 {\it Globale L\"osungen von gew\"ohnlichen linearen Differentialgleichungen mit zwei stark sigul\"aren Stellen}, Doctoral dissertation, University of Heidelberg
    \nonum Naundorf F 1976 {\it SIAM J. Math. Anal.} {\bf 7} 157

\bibitem{gom1} G\'omez F J and Sesma J 2007 {\it J. Comput. Appl. Math.} {\bf 207} 291

\bibitem{gom2} G\'omez F J and Sesma J 2010 {\it J. Phys. A: Math. Theor.} {\bf 43} 385302
    \nonum G\'omez F J and Sesma J 2012 {\it Eur. Phys. J. D} {\bf 66} 6
    \nonum Sesma J 2013 {\it J. Math. Chem.} {\bf 51} 1881

\bibitem{hard} Hardy G H 1956 {\it Divergent Series} (Oxford: Clarendon Press)

\bibitem{bedo} Bender C M and Boettcher S 1998 {\it Phys. Rev. Lett.} {\bf 80} 5243
    \nonum Dorey P, Dunning C and Tateo R 2001 {\it J. Phys. A: Math. Gen.} {\bf 34} 5679

\bibitem{mois} Moiseyev N, Certain P R and Weinhold F 1978 {\it Mol. Phys.} {\bf 36} 1613
    \nonum Moiseyev N and Friedland S 1980 {\it J. Comput. Phys.} {\bf 22} 1621
    \nonum Moiseyev N 1998 {\it Phys. Rep.} {\bf 302} 211

\bibitem{nobl} Noble J H, Lubasch M and Jentschura U D 2013 {\it Eur. Phys. J. Plus} {\bf 128} 93

\bibitem{nuss} Nussenzveig H M 1972 {\it Causality and Dispersion Relations} (New York: Academic Press)

\bibitem{gom3} G\'omez F J, Ra\~nada M F and Sesma J 1986 {\it J. Math. Phys.} {\bf 27} 552

\bibitem{ferr} Ferreira E M and Sesma J 2012 {\it J. Phys. A: Math. Theor.} {\bf 45} 415302

\bibitem{garc} Garc\'{\i}a-Calder\'on G and Peierls R 1976 {\it Nucl. Phys.} {\bf A265} 443

\end{thebibliography}
\end{document}